\documentclass[twocolumn,superscriptaddress,
	floatfix,amsmath,amssymb,nofootinbib]{revtex4-1}
\usepackage[english]{babel}
\usepackage{graphicx}
\usepackage{dcolumn}
\usepackage{bm}
\usepackage{verbatim}
\usepackage{mathrsfs}
\usepackage{natbib}
\usepackage{color}
\usepackage{hyperref}
\usepackage[dvipsnames]{xcolor}

\newcommand{\ket}[1]{\left| {#1} \right\rangle}
\newcommand{\bra}[1]{\left\langle {#1} \right|}
\newcommand{\ematrix}[3]{\left\langle {#1} \left|{#2}\right|{#3}\right\rangle}

\newcommand{\tr}{\operatorname{Tr}}

\newcommand{\tcg}{\textcolor[rgb]{0,.6,0}}

\newcommand{\be}{\begin{equation}}
\newcommand{\ee}{\end{equation}}

\begin{document}

\title{The equivalence principle and QFT:\\ Can a particle detector tell if we live inside a hollow shell?}
\author{Keith K. Ng}
\affiliation{Department of Physics and Astronomy, University of Waterloo, Waterloo, ON, N2L 3G1, Canada}
\author{Robert B. Mann}
\affiliation{Department of Physics and Astronomy, University of Waterloo, Waterloo, ON, N2L 3G1, Canada}
\affiliation{Institute for Quantum Computing, University of Waterloo, Waterloo, Ontario, N2L 3G1, Canada}
\affiliation{Perimeter Institute for Theoretical Physics, Waterloo, ON, N2L 2Y5, Canada}
\author{Eduardo Mart\'{i}n-Mart\'{i}nez}
\affiliation{Department of Applied Mathematics, University of Waterloo, Waterloo, ON, N2L 3G1, Canada}
\affiliation{Institute for Quantum Computing, University of Waterloo, Waterloo, Ontario, N2L 3G1, Canada}
\affiliation{Perimeter Institute for Theoretical Physics, Waterloo, ON, N2L 2Y5, Canada}

\begin{abstract}
We show that a particle detector can distinguish the interior of a hollow shell from flat space for switching times much shorter than the light-crossing time of the shell, even though the local metrics are indistinguishable. This shows that a particle detector can read out information about the non-local structure of spacetime even when switched on for scales much shorter than the characteristic scale of the non-locality. 
\end{abstract}

\maketitle

\section{Introduction}

It is well known that the vacuum state of a quantum field contains information about the structure of spacetime. For example, this is seen in the phenomenon of vacuum entanglement harvesting 
\cite{Valentini1991,Rez03,RRS05,Salton:2014jaa,Pozas2015,Pozas2016}.  
  But can  local measurements can provide non-local information about spacetime?  Suggestive evidence along these lines was recently obtained when it was  shown that the transition rate of particle detectors can be sensitive to the topology of spacetime, even when this topology is hidden behind an event horizon \cite{Smith:2013zqa}.  Furthermore, if a preferred direction is induced by spatial topology, this direction may be inferred from the dependence of correlations between the two detectors on their orientation \cite{Martin-Martinez:2015qwa}. Even in the simple case of the Einstein cylinder, the dynamics of the 
detectors exhibit beating behavior with no steady state established at late times, which distinguish from the results in noncompact flat spacetimes \cite{cilinder}.

In this paper we address the following problem:  is data from a  pointlike switchable antenna 
(that we take to be an Unruh-DeWitt detector) placed at the centre of a 
 massive hollow shell (not containing any matter) sufficient to distinguish the spacetime background from globally
flat spacetime?  Since the inside of the shell is locally flat, this task can only be accomplished 
 in classical terms  by sending a signal from an antenna to the shell (which could consist, for example, of the perturbation caused in the classical electromagnetic field due to switching on the antenna)  and wait for the emitted signal to bounce off the shell (or off the spacetime curvature generated by it outside of the shell if it is transparent to the emitted radiation) and return to the antenna. The echo carries information about the shell.  Consequently, if the field is in the ground state, it is impossible to use any observable of the antenna to determine whether or not it is located inside a hollow shell until waiting at least the light-crossing time of the shell.

 In quantum theory, however, the vacuum has information about `boundary conditions'; even a transparent shell, if  massive, will curve spacetime. So, at least in principle, the vacuum state of the field does have information about the global structure of spacetime. It is not clear, however, if a local detector would be able to read out that information if it is only allowed to interact with the field on timescales much smaller than the light-crossing time of the shell. One would expect that, to extract non-local information about the spacetime background, it would have to be `switched on' at least for timescales comparable to the light-crossing time of the shell. 

We show here that, perhaps unexpectedly, an antenna interacting quasi-locally with the vacuum state of a quantum field inside a hollow massive transparent shell is able to distinguish its location from that of a globally flat Minkowski spacetime  for timescales much smaller than the light-crossing time of the shell.
More technically, suppose that an observer is placed in one of two possible spacetimes, $\mathcal{M}_1$ and $\mathcal{M}_2$. If the metric near the observer (i.e. within a neighbourhood $\mathcal{U}$) is identical in both spacetimes, the Einstein equivalence principle states that the two spacetimes should be indistinguishable with respect to any non-gravitational local (classical) experiment \cite{eep}. The term \textit{local} here indicates that we should not allow any interaction with an external field. One natural way to relax the assumptions is to permit interaction with an external field, but have the field in $\mathcal{U}$ in both cases in the vacuum state. As discussed above, classically, this still does not permit us to distinguish the two cases since the ground state of a classical theory (zero field amplitude) contains no non-local information about the structure of spacetime: as long as we are limited to `local' measurements, i.e. one whose causal diamond lies inside $\mathcal{U}$, any experiment will still produce the same result, since the local background is the same. In other wor\text{d}s, classical measurements cannot distinguish the two spacetimes `faster than light'.
 
The situation drastically changes once we consider quantum fiel\text{d}s. It is well-known that, as opposed to the classical ground state, the vacuum state can carry non-local information about the boundary conditions: one of the most famous examples is the Casimir effect \cite{casimir1948}, in which two parallel uncharged conductive plates in a vacuum are attracted to each other. In that case, it might be said that even if there is no real photon exchange between the plates, the field caries information of the presence of one conductive plate to another. For example, the fact that there is no need to exchange real photons to signal has been used to propose novel communication setups where information can be sent without energy being exchanged between sender and receiver \cite{Comm2,Blasco2015,Blasco2016}.


In this paper we demonstrate that -- even with a non-interacting (i.e. transparent) shell with no net effect on the local gravitational field around an Unruh-DeWitt detector --  the detector can still determine the presence of the shell through the non-local effects of gravity on the field vacuum. This fact can be used to distinguish local flatness from global flatness.  Thus, we wish to show  not only that the vacuum state of an external field carries non-local information about the \textit{gravitational} field,  but also that a detector can read out that information locally.

\section{Basic Setup}
\label{sec2}

Consider an Unruh-DeWitt detector \cite{dewitt1980} in curved space that can be switched on and off. 
Although simple, this model captures the fundamental features of the light-matter interaction \cite{Wavepackets,Alvaro,Pozas2016}. 
We can represent the detector interaction of the detector with a scalar field in space using the following interaction Hamiltonian:
\begin{equation}\label{Hamiloquai}
\hat{H}_{I}(\tau)=\lambda\chi(\tau)\hat{\mu}(\tau)\hat{\phi}(x(\tau)).
\end{equation}
In the above, $\chi(\tau)$ is the switching function, $\hat{\mu}(\tau) = {e^{i\hat{H}_{0,d} \tau}\hat{\mu}(0)e^{-i\hat{H}_{0,d} \tau}}=e^{-i\Omega \tau}\hat{\sigma}^+ + e^{i\Omega \tau}\hat{\sigma}^-$ is the interaction picture monopole operator  of the detector  with $\hat{H}_{0,d}$ the Hamiltonian of the free detector, and $\hat{\phi}(x)$ is the field operator. We may then calculate the response of the detector to the ambient scalar field; since the curvature of space affects the scalar field modes, the response of the detector is sensitive to the shape of spacetime. However, it is \textit{a priori} unclear how the detector's behaviour depen\text{d}s on far-away features. In particular, inspired by the Casimir-Polder effect \cite{CasimirPolder,Hu2004,Alvaro} and the Quantum-Collect Calling results \cite{Comm2}, we ask the following question: if the detector is located in a flat region, can we determine whether we are within a shell of matter  (as opposed to empty space) by measuring its response? To find a signature of the detector's response of the presence of the shell, we shall explore how the transition probability of the antenna depen\text{d}s on the presence of the matter shell.

Let us be more concrete. Suppose that we begin with the detector in its ground state and the field in the vacuum. We then modulate the switching function $\chi(\tau)$ from zero to a finite value from some time $\tau=\tau_i$ to $\tau_f$; this finite-time interaction results in a nonzero probability of detector excitation. The excitation probability of the detector can be represented as the expected value of the state projector $P_1=|1\rangle_d\langle 1 |$ after the switching is complete at time $\tau_f$,  when  $\chi(\tau_f)=0$. 
  
Using $\hat{\rho}$ to indicate the state of the total system, and $\hat{P}_1$ to represent the (projector to the) excited state of the detector, we can write the transition probability as
\begin{equation}
P=\tr \left(\hat{P}_1 \hat{\rho}(\tau_f)\right).
\end{equation}
We can then ask  for a detector placed at the centre of a shell of matter  whether this probability  differs  from that obtained in the absence of the shell and, if any, how precisely this difference 
can be computed. The transition probability, to leading order, is well-known \citep{birrelldavies}:
\begin{align}
P\approx&\tr\left(\hat{P}_1(\tau_f)\hat{U}_1\hat{\rho}(0)\hat{U}_1^\dagger\right)\nonumber\\
=&\int_{-\infty}^{\infty}\text{d}\tau_1\int_{-\infty}^{\infty}\text{d}\tau_2\tr\left(\hat{P}_1(\tau_f)\hat{H}_I(\tau_1)\hat{\rho}(0)\hat{H}_I(\tau_2)\right)\nonumber\\
=&\lambda^2\left|{ _d}\negthinspace\ematrix{\Omega}{\hat{\mu}(0)}{0}_d\right|^2 \mathcal{F}(\Omega),
\end{align}
where $\Omega$ is the energy gap of the detector.
We write this separation of $\mathcal{F}(\Omega)$ from the detector part in order to emphasize that this part is independent of the properties of the detector; this ``response function'' depen\text{d}s only on the properties of the field. 
It can be expressed as
\begin{align}
\mathcal{F}(\Omega)=&\int_{-\infty}^{\infty}\text{d}\tau_1\int_{-\infty}^{\infty}\text{d}\tau_2\, e^{-i\Omega(\tau_2-\tau_1)}\chi(\tau_1)\chi(\tau_2)\nonumber\\
&\times {}_f\negthinspace\bra{0}\hat{\phi}(x(\tau_2))
\hat{\phi}(x(\tau_1))\ket{0}_f.\nonumber\\
\end{align}
Note that here and afterwar\text{d}s, $\Omega$ is the detector gap proper frequency, i.e., at the detector location. This allows us to compare different spacetimes, and is the quantity an experimenter at the detector would control.

At this point, let us consider a static spherically-symmetric spacetime, whose  line element is
\begin{equation}\label{mtrc}
\text{d}s^2 = -\alpha^2 \text{d}t^2 + a^2 \text{d}r^2 + r^2 \text{d}\Omega^2
\end{equation}
where $\text{d}\Omega_2^2$ is the usual line element of the 2-sphere. Let us then consider the specific case of the massless scalar field obeying the Klein-Gordon equation
\begin{equation}
\square \Psi \equiv \nabla^\mu \nabla_\mu \Psi= 0 \; .
\end{equation}

To simplify further, let us write these operators in the fixed frequency mode  basis with respect to some quantization time $t$.
Recall that we can write the field operators as
\begin{align}
 \hat{\phi} &(x(\tau))=\nonumber\\
&\sum_{l,m}\int_0^\infty \text{d}\Omega \, \left(\hat{a}_{\omega lm}\Psi_{\omega lm}(x(\tau))
+ \hat{a}^\dagger_{\omega lm}\Psi^\dagger_{\omega lm}(x(\tau)) \right) 
\end{align}
where $\Psi_{\omega lm}(x)$ is the solution to the Klein-Gordon equation with energy $\omega$ and angular momentum numbers $l,m$  and $a_{\omega l m}$ is the corresponding mode annihilator, with the usual commutation relations,
$$
[\hat{a}_{\omega' l'm'},\hat{a}^\dagger_{\omega lm}]=\delta(\omega'-\omega)\delta_{l'l}\delta_{m'm}
$$
In this basis, we can rewrite the response function as
\begin{align}
\mathcal{F}(\Omega)=&\int_{-\infty}^{\infty}\text{d}\tau_1\int_{-\infty}^{\infty}\text{d}\tau_2\, e^{-i\Omega(\tau_2-\tau_1)}\chi(\tau_1)\chi(\tau_2)\nonumber\\
&\times\sum_{lm}\int_0^\infty \text{d}\Omega\,\Psi_{\omega lm}(x(\tau_2))\Psi_{\omega lm}^\dagger(x(\tau_1))
\label{responsefunction}
\end{align}

Note that while the detector gap $\Omega$ is a proper frequency at the detector, the basis modes $\Psi_{\omega lm}$ are labelled by their frequency with respect to $t$ instead, as this $\omega$ is the quantity that appears in the Klein-Gordon equation.

\section{Solving the Klein-Gordon equation}

Rewriting the action of the  d'Alembertian ($\square$) on the scalar gives
\begin{equation}
\square\Psi=\frac{1}{\sqrt{-g}}\partial_\mu(\sqrt{-g}g^{\mu\nu}\partial_\nu \Psi),
\end{equation}
where $g= -\alpha^2a^2r^4\sin^2\theta$ is the determinant of the metric \eqref{mtrc}.   The Klein-Gordon equation becomes
\begin{align}
0=&-\frac{1}{\alpha^2}\partial_t^2\Psi + \frac{1}{\alpha a r^2}\partial_r\left(\frac{\alpha}{a}r^2\partial_r\Psi\right) \nonumber\\
&+\frac{1}{r^2 \sin\theta}\partial_\theta\left(\sin\theta \,\partial_\theta\Psi\right)+ \frac{1}{r^2 \sin^2 \theta}\partial^2_\phi\Psi
\end{align}
assuming the metric is static, i.e. $\partial_t \alpha = \partial_t a = 0$. Noting that  the angular coordinates and the time coordinate separate, we  write
\begin{equation}
\Psi=\frac{1}{\sqrt{4\pi \omega}}e^{-i\omega t}\psi_{\omega l}(r)Y_{lm}(\theta,\phi),
\label{sepvars}
\end{equation}
and obtain the suggestive form
\be
0=\omega^2\psi + \frac{\alpha}{ a r^2}\partial_r\left(\frac{\alpha}{a}r^2\partial_r\psi\right) 
-\alpha^2\frac{l(l+1)}{r^2}\psi
\ee
upon suppressing  the subscripts and the $r$ dependence of $\psi$ for brevity. Further defining $r^*$ such that $\partial_{r^*}=\frac{\alpha}{a}\partial_r$  and  setting $\psi=\rho/r$ yields
\begin{align}
0=\omega^2\rho + \frac{1}{r}\partial_{r^*}\left(r^2\partial_{r^*}\left(\frac{\rho}{r}\right)\right) -\alpha^2\frac{l(l+1)}{r^2}\rho
\label{radialeqnrough}
\end{align}
or
\begin{align}
0=\partial_{r^*}^2\rho+\left(\omega^2-V_{l}(r)\right)\rho 
\label{radialeqn}
\end{align}
where the effective potential is
\begin{equation}
V_{l}(r)=\alpha^2\frac{l(l+1)}{r^2}+\frac{1}{r}\frac{\alpha}{a}\partial_{r} \frac{\alpha}{a}
\label{diffeffpot}
\end{equation}
and we note that the effective potential is a function of the original radial coordinate $r=r(r^*)$.

 Let us now consider the shell spacetime. Suppose the shell's radius is $R$, and the ADM mass is $M$. We can then write the line element as
\begin{equation}
\text{d}s^2=
\begin{cases}
-f(r)\text{d}t^2+\frac{1}{f(r)}\text{d}r^2+r^2\text{d}\Omega^2, &r>R\\
-f(R)\text{d}t^2+\text{d}r^2+r^2\text{d}\Omega^2, &r<R,
\end{cases}
\end{equation}
where $f(r)=1-2M/r$. (Strictly speaking, one should use different variables for $r$ inside and outside the shell, and then smoothly match the radial metrics; however, our treatment of the differential equation handles this later.) Notice first that the time component of the metric, $g_{tt}$, is continuous; this is in accordance with the first Darmois condition \cite{darmois,israel1966},  that the induced metric on the shell should be continuous. If we assume our shell is made of a perfect fluid, we can then apply the Lanczos equation \cite{israel1966} to find that the surface density and surface pressure of our shell are
\begin{align}
\sigma=&\frac{1}{4\pi R}\left(1-\sqrt{1-\frac{2M}{R}}\right),\\
p=&\frac{1-M/R-\sqrt{1-2M/R}}{8\pi R \sqrt{1-2M/R}}.
\end{align}
As  expected, the pressure is positive, and the quantities are defined only for $R>2M$. Note that $\sigma$ is not equal to $M/4\pi R^2$; the difference is due to binding energy.

This metric is quite familiar on either side of the shell: inside, it is just Minkowski spacetime with a constant scaling factor on the time coordinate, while outside it is the Schwarzschild metric. As a result,  the effective potential becomes
\begin{equation}
V_l(r)=
\begin{cases}
f(r)\left(\frac{l(l+1)}{r^2}+\frac{2M}{r^3}\right),&r>R\\
f(R)\left(\frac{l(l+1)}{r^2}\right),&r<R.
\end{cases}
\end{equation}
To complete the description of the scalar field, we need to determine what the Klein-Gordon equation looks like \textit{on} the shell. This is troublesome, since $a$ is discontinuous, and therefore $\alpha/a$ is discontinuous. This implies that the effective potential \eqref{diffeffpot} is discontinuous, and so we must proceed with caution.

We now solve the differential equation weakly.
In order to find weak solutions, let us integrate equation \eqref{radialeqnrough} with respect to $r^*$ on an infinitesimal interval near the shell. Clearly, the only singular term of that equation is the term involving $\partial_{r^*}$. Employing the common notation
\begin{equation}
[X]=\lim_{r\rightarrow R^+} X - \lim_{r\rightarrow R^-} X;
\end{equation}
 we quickly find $0=[r^2 \partial_{r^*}(\psi)]$, or  more succinctly,
\begin{equation}
0=\left[\partial_{r^*}\psi\right] =\left[\frac{\alpha}{a}\partial_r \psi\right]
\label{discontinuity}
\end{equation}
showing that in $r^*$ coordinates, any weak solution $\psi$ to the Klein-Gordon equation has a continuous derivative across the shell. However, the derivative of $\rho=r\psi$ is discontinuous, so this must be kept in mind. Furthermore, while the derivative with respect to the original coordinates is discontinuous, the nature of the discontinuity is independent of the energy and other quantum numbers of the mode.

\section{Inside and Outside}

In any case,  equation \eqref{discontinuity}, along with the continuity of $\rho$ across the shell, allows us to connect Minkowski-space modes inside the shell to Schwarzschild-space modes outside. Notably, since $\alpha=\sqrt{f(R)}$ and $a$ are constant inside the shell, it is simplest to have $r=r^*=0$ at the center;  we then have $r=\alpha r^*/a=\alpha r^*$ (as $a=1$). We can quickly simplify  equation \eqref{radialeqnrough} to
\be 
0 = \tilde{r}^2\psi + 2\tilde{r}\partial_{\tilde{r}}\psi + \tilde{r}^2\partial_{\tilde{r}}^2 \psi - l(l+1)\psi.
\ee
which  is just the spherical Bessel equation, with $\tilde{r}=\omega r^*$, and thus the inner solutions are (up to normalization) $\psi_{  \omega l} \sim j_l(\omega r^*)=j_l(\tilde{\omega} r)$ where $\tilde{\omega}=\omega/\sqrt{f(R)}$.

While at first glance this appears to imply that the inside solution is independent from the outside, this is not the case: if we wish to preserve normalization of the modes, we must have
\begin{equation}
\int_0^\infty \text{d}r^* \rho_{\omega_1 l_1}(r) \rho_{\omega_2 l_2}(r)=2\pi \delta(\omega_1-\omega_2) \delta_{l_1,l_2}
\end{equation}
where $ \rho_{\omega l}(r) = r \psi_{\omega l}$.
In order for this to occur, since $r^* \rightarrow r$
as $r\rightarrow \infty$, the asymptotic behaviour of the modes must be $\rho_{\omega l}(r)\rightarrow 2\sin(\omega r + \theta)$ as $r\rightarrow \infty$ for some phase $\theta$. Therefore, the position and mass of the shell affects the `transmission' of the mode through the shell, and therefore the normalization of the mode inside the shell. In the absence of such a shell, since the asymptotic behaviour of the spherical Bessel functions is $j_l(\omega r)\rightarrow \sin (\omega r + \theta)/(\omega r)$, we must multiply the spherical Bessel function by $2\omega$; thus $\psi = 2\omega j_l(\tilde{\omega} r)$ under this normalization scheme. 

We can combine the inner solution with the discontinuity equation \eqref{discontinuity} in order to find the normalization factor $A_{\omega l}$. To do this, we begin with the unnormalized inner solution $\tilde{\psi}=j_l(\tilde{\omega}r)$, and find its derivative across the shell with respect to $r_*$. Since $\partial_{r^*}=\frac{\alpha}{a}\partial_r$, that means the derivative is 
$\partial_{r^*}\tilde{\psi}|_{r=R} =\frac{\alpha}{a}j'_l(\tilde{\omega}R)=\tilde{\omega} \sqrt{f(R)}j'_l(\tilde{\omega}R)$. We can then input the value of $\tilde{\psi}$ and its derivative into the outer differential equation and integrate outwar\text{d}s; then, as we approach infinity,  equation \eqref{radialeqn} implies $\tilde{\psi}\rightarrow 2A_{\omega l}^{-1}\sin(\tilde{\omega}r^*)/r^*$, and so $\psi = A_{\omega l}\tilde{\psi}.$ In particular, for $r=0$, $\psi = \delta_{l,0}A_{\omega l}.$

As a side note, if we wish to represent the mode in terms of $\rho$ rather than $\psi$, this result shows that as $r\rightarrow 0$, we must have $\rho/r \rightarrow \delta_{l,0}A_{\omega l}$; we only have a singularity because our transformation misbehaves there.  If we wish to interpret the radial equation as a scattering problem, this lea\text{d}s to unusual consequences---as energy goes to infinity,  the effect of the discontinuity does not diminish even though the scattering due to the potential vanishes;  it is thus  the largest difference from the flat case. Specifically, while $\rho$ is continuous across the shell,  its derivative with respect to the tortoise coordinate experiences a jump of
\begin{align}
[\partial_{r^*}\rho]=&[\partial_{r^*}(r)]\rho\nonumber\\
=&\left(f(R)-\sqrt{f(R)}\right)\rho \quad .
\label{shelljump}
\end{align}
In other wor\text{d}s, the value of $\partial_{r^*}\rho$ just outside the shell is $\left(f(R)-\sqrt{f(R)}\right)\rho$ greater than the value immediately inside.
Notably, since this contribution is proportional to $\rho(R)$, a resonant effect is present.  We emphasize that the discrete nature of the shell is \textit{not} essential for this phenomenon: while a shell of finite thickness would not exhibit a discontinuity \textit{per se}, a similar relation between $\rho$ and $\partial_{r^*}\rho$ would  exist on the shell, as it can be derived from integration of the Klein-Gordon equation.

\section{Static Detector}

In the special case of the static detector, equation \eqref{responsefunction}  is tremendously simplified. In that case, we can assume (by spherical symmetry) that $\theta=\pi/2, \phi=0$, and so we can write
\begin{align}
&\mathcal{F}(\Omega)=\int_{-\infty}^{\infty}\text{d}\tau_1\int_{-\infty}^{\infty}\text{d}\tau_2\, e^{-i\Omega(\tau_2-\tau_1)}\chi(\tau_1)\chi(\tau_2)\nonumber\\
&\times\sum_l\int_0^\infty \text{d}\Omega\,e^{-i\tilde{\omega}(\tau_2-\tau_1)}
\left|\sqrt{\frac{2l+1}{16\pi^2\omega}}\frac{\rho_{\omega l}(r)}{r}\right|^2
\end{align}
since $t=\tau/\alpha$.   
Given a sufficiently smooth switching function, we can swap the order of integration, and combine the exponentials. 
The resulting integrals simply describe the Fourier transforms of the switching function. Using the unitary definition of the Fourier transform,
$$\hat{\chi}(\Omega)=\frac{1}{\sqrt{2\pi}}\int_{-\infty}^\infty d\tau\,e^{-i\Omega\tau}\chi(\tau),$$
we end up with
\begin{align}
\mathcal{F}(\Omega)=\sum_l\int_0^\infty \text{d}\Omega\,
\hat{\chi}(-\Omega-\tilde{\omega})\hat{\chi}(\Omega+\tilde{\omega}) \frac{2l+1}{8\pi\omega r^2}\left|\rho_{\omega l}(r)\right|^2.
\end{align}
Finally,   $\hat{\chi}(\Omega)=\bar{\hat{\chi}}(-\Omega)$ since the switching function is real. Rewriting the integral in terms of $\tilde{\omega}$, we get the simple expression
\be 
\mathcal{F}(\Omega)=\sum_l\int_0^\infty d\tilde{\omega}\, |\hat{\chi}(\Omega+\tilde{\omega})|^2 
\frac{2l+1}{8\pi\tilde{\omega} r^2}\left|\rho_{\omega l}(r)\right|^2
\label{F(r)}
\ee
and we note that since the factor $\text{d}\Omega/\omega$ appears in the integral, the change in variables does not directly introduce $\alpha(r)$ into the expression. The only actual dependence is through the mode function, which depen\text{d}s on $\omega=\alpha(r)\tilde{\omega}$; in fact, if we choose to express the mode in terms of the proper time at the detector (rather than the coordinate time), the `local energy' becomes $\tilde{\omega}$, and thus the expression can be written without direct reference to $\omega$.

The appearance of the Fourier transform of the switching function in this expression has a number of unexpected consequences. The most important is that by suitably varying the switching function, we can `focus' on a single mode energy $\omega$. This is especially relevant in the special case where $r=0$, since we can then eliminate all modes with $l\neq 0$. We then find
\begin{align}
\mathcal{F}(\Omega)=&\int_0^\infty d\tilde{\omega}\,\frac{1}{8\pi\tilde{\omega}}
 |\hat{\chi}(\Omega+\tilde{\omega})|^2\left|A_{\omega 0}\right|^2
\label{F(0)}
\end{align}
in this limit.

Equation \eqref{F(0)} is the pivotal equation for our paper. We 
 note that it hol\text{d}s even when the spacetime is Minkowski, with the substitution $A^M_{\omega 0}=2\omega=2\tilde{\omega}$ (since in that case $f(R)=1$). Consequently we can use this final expression to determine whether or not the detector is in a shell, even if the shell is far from the detector.  In particular, if for \textit{any}  frequency $\left|A_{\omega 0}\right|\neq 2\tilde{\omega}$, we can conclude that a suitably selected switching function can distinguish the two possibilities.  It is easy to show that the adiabatic switching case correspon\text{d}s to $\tilde\chi(\omega)\propto\delta(\omega)$, and thus represents the `ideal' case; we can then simply tune the gap to the desired frequency.
 
 Perhaps more surprisingly, eq. \eqref{F(0)} shows that a detector  is in principle capable of distinguishing the two different backgroun\text{d}s, even if it remains turned on for a very short time. To study how a detector can so discriminate, we need to assess how much frequency resolution is necessary to perform this task.  The higher the frequency resolution required, the more time the measurement requires. If this time is less than the light-crossing time of the shell, we can say  a local measurement is sensitive to the presence of the shell, since signals from the detector do not have time to reach the shell and return.  Conversely, if the required frequency resolution is too fine, then no `local' measurement can detect the shell.

In the particular case of the transparent spherical shell, we found that the primary difference between $A_{\omega 0}$ and $2\tilde{\omega}$ was the `resonant' effect of the shell, as governed by equation \eqref{shelljump}. Since the corresponding term appearing in \eqref{F(0)} is $|A_{\omega 0}|^2$, it is fairly  simple to show that the shell resonance has a `period' of approximately $\pi/R$ in $\omega$, which correspon\text{d}s via the energy-time uncertainty relation to a (coordinate) temporal width of $R/2\pi$, on the order of the shell-crossing time. However, even if the frequency resolution of our switching function is lower, i.e. its temporal width is smaller than $R/2\pi$, it is still possible to observe a (smaller) signal. 

 We therefore have a means of quantum-mechanically distinguishing the gravitational field inside an empty shell or cavity from that of flat space-time, a feat that is classically impossible.  Furthermore,
 equations \eqref{F(r)} and \eqref{F(0)} are extremely general; we can consider almost any spherically symmetric spacetime and get a similar result. The idea of varying the switching function to focus on a single energy means that, in general, the detector can {extract from the vacuum} the properties of spacetime far from its location, given sufficient time.

%

\section{Numerical computation and results}

 In order to demonstrate that `local'  determination of curvature is possible, we evaluate the expression \eqref{F(0)} for Minkowski and shell spacetimes with a Gaussian switching function, with temporal width much shorter than the light-crossing time of the shell. Specifically, we use
\begin{equation}
\chi(\tau)=e^{-\tau^2/2\sigma^2}. \label{GaussSw}
\end{equation}
Classically, we would expect that the minimum time to distinguish the two scenarios is the time for a signal to travel from the detector to the shell and back; thus,  for a local measurement, we require that $\sigma \ll 2R$. While one might argue that the `tails' of the Gaussian still allow for classical communication, the value of the switching function far into the tails is extremely small; we shall discuss  this point below.

\begin{figure}[hbtp]
\centering
\includegraphics[width=\columnwidth]{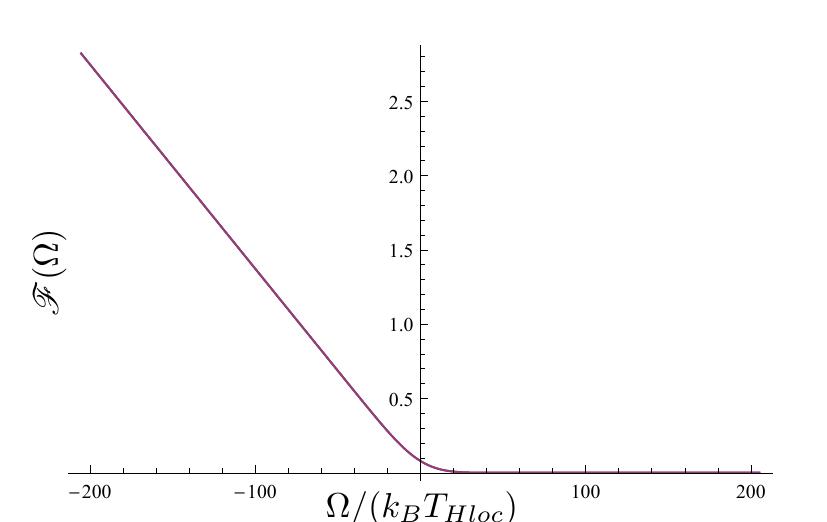}
\caption{Response function for a detector in a shell (blue) vs. flat space (red) for varying detector gaps}
\label{Fgraph}
\end{figure}

For purposes of demonstration,  we compute all quantities in units {where $\hbar=c=R_{\textrm{Sch}}=1$}. $R_{Sch}$ is the
(hypothetical) Schwarzschild radius of the shell. This will set  the mass of the shell, and we will vary its radius; we will initially take it to be $R=3R_{\textrm{Sch}}$,  equal to the innermost stable circular orbit. We will use the local Hawking energy of an equal mass black hole, i.e. $k_B T_{Hloc},$ to measure energies.
 We select $\sigma = 0.5$, which is well below the crossing time of the shell. In order to save computational time, we first compute the values of the radial modes at the center for a mesh of values and interpolate. We then compute the response function \eqref{F(0)} (where a negative gap indicates that the detector is initialized in its excited state), and compare the results for a detector in a shell (in blue) to a detector in flat space (in red). The results are shown in Fig. \ref{Fgraph}.

While the curves in Fig. \ref{Fgraph} are very similar, they are not identical and in fact can
be distinguished. We plot the difference between responses, $\mathcal{F}_{shell}(\Omega)-\mathcal{F}_{flat}(\Omega)$, in Fig. \ref{diffFgraph} below.
\begin{figure}[hbtp]
\centering
\includegraphics[width=\columnwidth]{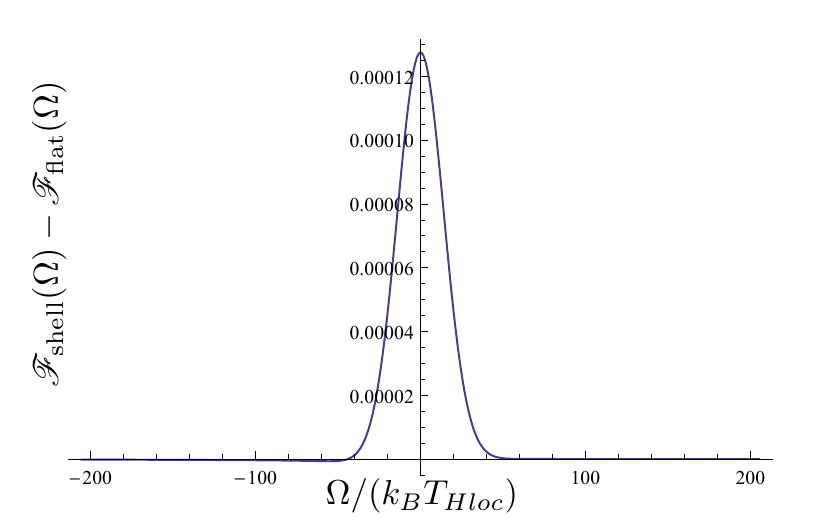}
\caption{Absolute difference in response function between a detector in a shell vs. in flat space, for varying detector gaps.}
\label{diffFgraph}
\end{figure}
As we can see, the largest absolute difference is visible at the smallest values of the gap. It is also interesting to see that the absolute difference is somewhat symmetric in gap, since the expressions for $\mathcal{F}(\Omega)$ given in \eqref{F(0)} are manifestly asymmetric.

\begin{figure}[hbtp]
\centering
\includegraphics[width=\columnwidth]{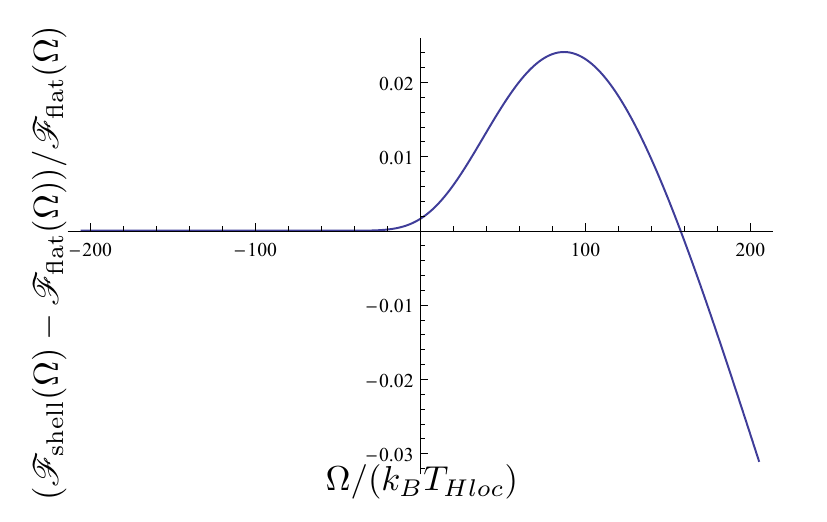}
\caption{Relative difference in response function between a detector in a shell vs. in flat space, for varying detector gaps.}
\label{relFgraph}
\end{figure}

However, the absolute difference in response functions may not be a good measure of how distinguishable they are. Operationally, a better measure of distinguishability is the \textit{relative} difference between the two responses, i.e. $(\mathcal{F}_{shell}(\Omega)-\mathcal{F}_{flat}(\Omega))/\mathcal{F}_{flat}(\Omega)$, shown in Fig. \ref{relFgraph}.
 {
In this case, we see that while the absolute difference in response is approximately symmetric, the decay in the ``background'' response causes a larger relative difference when the gap is positive. The difference is within a single order of magnitude, however; while the positive gap detector is more sensitive to the difference in response, it is not excessively so. 
 A local maximum may also be observed at around $\Omega=90k_B T_{Hloc}$. In other words, while the upper energy scale of the graph is larger than is usually considered in the black hole case, it is still quite small compared to the mass-energy of the shell.

We see that the strongest relative differences between geometries can be observed at large \textit{positive} gaps, even though the absolute value of the switching probability is very small there. This may be explained as follows: since the tails of the Gaussian switching function decay super-exponentially (in both the time and frequency domains), a large (positive) energy gap correspon\text{d}s to an extremely quickly decaying integrand; it can be shown that the modes differ significantly at low energy, as shown in Fig. \ref{modegraph}. These factors are a result of our integration over positive frequencies only; they would not come into play if we had to integrate over negative $\omega$, i.e. if the `vacuum' state had non-zero particle content.  As well, it is not clear whether this phenomenon is specific to this order of perturbation: it is possible that inclusion of higher order terms would eliminate this anomaly. {In fact, we have also found that a more slowly decaying Fourier-transformed switching function \textit{does} remove this effect, since it appears to depend on the super-exponential decay of the Gaussian. For instance, the switching function $\chi_2 (\tau)=(1+(\tau/\sigma)^2)^{-1}$ Fourier transforms into a Lorentzian, and thus the relative response is constant for \textit{any} positive gap.}
\begin{figure}[hbtp]
\centering
\includegraphics[width=\columnwidth]{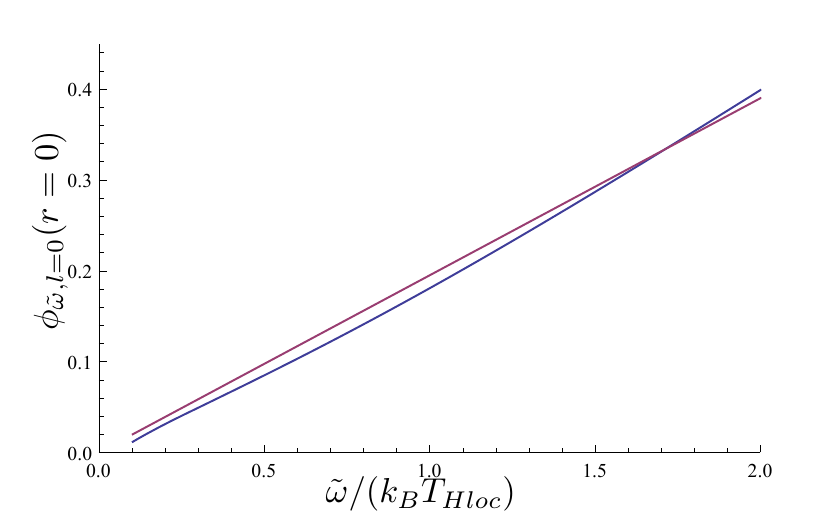}
\caption{Value of modes in shell (blue) vs. flat space (red) at origin}
\label{modegraph}
\end{figure}

{
In any case, it is quite likely that the relevance of this unusual phenomenon is limited: since the \textit{absolute} transition rate is very small at the relevant values of the gap (e.g. $\mathscr{F}(\Omega=20)<10^{-45}$), and because of all the previously mentioned caveats, it is likely that any experimenter would find this ``high-signal'' regime inaccessible in practice. A better strategy would be to use a detector with a small positive gap, which balances the need for a detectable absolute difference in response with the need for a small background response.}

As a secondary check on our calculation, we now plot in Fig. \ref{rsgraph} the switching energy as a function of the shell radius, for fixed switching parameters  $\sigma=0.2$, $\Omega=0$.  Note that with this choice of parameters, the switching timescale $0.2$ is much less than the shell-crossing time $R>1$, so we are comfortably within the `local' regime.
 These parameters correspond to a single switching energy in the flat case, which is shown on the graph below in red.  As expected, the larger the radius of the shell, the weaker the ability of the detector to distinguish the cavity interior from pure Minkowski spacetime.
\begin{figure}[hbtp]
\centering
\includegraphics[width=\columnwidth]{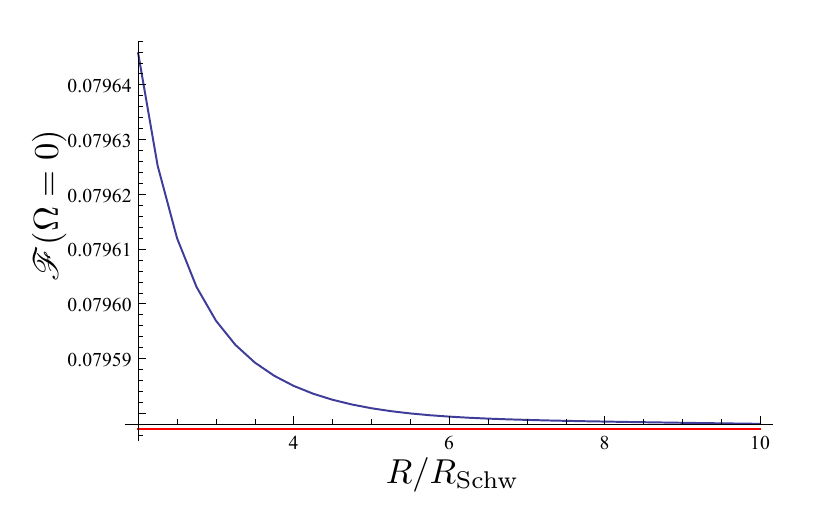}
\caption{Response function vs. shell radius. Value for flat case in red.}
\label{rsgraph}
\end{figure}

We now comment on the `classically communicating tails'. While \textit{a priori} the presence of these tails would seem to weaken our arguments, we chose a Gaussian with width an order of magnitude smaller than the crossing-time of the shell. This implies that the part of the response function contributed by these tails is exceedingly small, certainly orders of magnitude smaller than the order $10^{-4}$ differences observed in Fig. \ref{relFgraph}. Indeed, we have also conducted calculations where the switching function is modified to have finite support, i.e. removing the tails, with no observable difference in the results. However, we note here that while using a sudden switch to remove the tails is tempting, it introduces large tails in the frequency domain of the switching function, which makes numerical integration very troublesome; of course, sudden switching requires an infinite amount of energy, and this is its manifestation. While using the methods of Louko and Satz \cite{louko2007} to subtract out the singularities is possible, we feel that smooth switching allows for better interpretation; using smoothly varying switching functions of compact support yiel\text{d}s no observable difference. However, care is needed: the phenomenon of greater relative difference in responses at large $\Omega$ depen\text{d}s on the high-frequency behaviour of the tails, which becomes unphysical for sudden switching, and is generally sensitive to the precise way the detector is switched.

\section{Conclusion}

 We have shown that it is possible to distinguish a flat spacetime from the interior of a transparent (but masssive) shell with local particle detector measurements. i.e., measurements where the particle detectors are allowed to interact with the field only for timescales much shorter than the shell light-crossing time, and even when the state of teh field is the vacuum.

Concretely, the difference between a shell and a flat space \textit{can} be observed  locally,  even from inside. We demonstrated this numerically for a very compact shell with radius $R=3R_S$, and a detector with an interaction timescale much smaller than the shell's light-crossing time. We discussed the conditions under which the strongest effects may be observable.   We also evaluated the switching energy for various shell radii, confirming that the effect vanishes as the shell radius goes to infinity. Counterintuitively, a large positive energy gap appears to yield the best relative results. However,  the main question of this paper has been answered: it is indeed possible to   quantum-mechanically distinguish the inside of a shell from flat space in a time much smaller than the classical communication time from the detector to the shell, i.e. `locally'.

 One possible next step would be to determine whether more spatial information can be extracted: for instance, whether one could recover the density profile of a general shell, or perhaps non-radially symmetric objects. Another possibility would be to extend this work to more realistic detector models, i.e. of finite spatial extent, or the case of an hydrogen-like atom interacting with an electromagnetic field \cite{Pozas2016}.  We also may speculate on whether one can prove a lower bound on the time/energy required to distinguish different spacetimes; such a result would be of interest to those researching the firewall question \cite{amps,Braunstein2013}, for instance, in the spirit of work done in \cite{hotta2013}.  To this end, a computation for a detector at the centre of a collapsing shell would be of considerable interest.

\section*{Acknowledgements}
We would like to thank Shih-Yuin Lin for very helpful discussions and insights during the development of this paper.
This work was supported in part by the Natural Sciences and Engineering Research Council of Canada.

\appendix
\section{Normalization}
We wish to verify that the modes described here are normalized with respect to the Klein-Gordon inner product, which in curved space is
\begin{equation}
(\Psi_1,\Psi_2)=i\int_\Sigma \text{d}\sigma\, n^\mu (\Psi_1^* \nabla_\mu \Psi_2 - \Psi_2 \nabla_\mu \Psi_1^*)
\end{equation}
over some Cauchy surface $\Sigma$, with normal $n^\mu$.
Of course, while this is not positive definite for generic solutions of the Klein-Gordon equations, it will work for positive frequency solutions with respect to the Killing time. In this case, the obvious choice of Cauchy surface is a constant-$t$ surface; since our solutions are time-independent we are free to choose whichever we like. The normalized normal vector is then $(1/\alpha(r),0,0,0)$, and the surface area element is $\text{d}\sigma=\sqrt{a^2(r)r^4 \sin^2 \theta}$.

If we then write our modes as in \eqref{sepvars}, since the radial functions $\psi$ are real, we then find the inner product looks like
\begin{align*}
(\Psi_1,\Psi_2)=&-i\int_{\Sigma} \text{d}r \text{d}\theta \text{d}\phi \left(\frac{a(r)r^2 \sin \theta}{\alpha(r)}\right)\nonumber\\
&\times(\Psi_1^* \partial_t \Psi_2 - \Psi_1 \partial_t \Psi_2^*)\nonumber\\
=&\frac{\omega_1+\omega_2}{4\pi\sqrt{\omega_1\omega_2}}e^{i(\omega_2-\omega_1)t}\nonumber\\
&\int_{\Sigma} \text{d}r \,\text{d}\Omega^2\, \frac{a(r)r^2}{\alpha(r)}
\psi_1^*(r)\psi_2(r)Y^*_{l_1m_1}Y_{l_2m_2}.
\end{align*}

Rewriting in terms of $\rho=\psi/r$ and rearranging a few things gets us
\begin{align*}
(\Psi_1,\Psi_2)
=& \frac{\omega_1+\omega_2}{4\pi\sqrt{\omega_1\omega_2}}e^{-i(\omega_2-\omega_1)t}
\int_0^\infty \frac{a(r)\text{d}r}{\alpha(r)}\rho_1^*(r)\rho_2(r) \nonumber\\
&\int_{S^2}\text{d}\Omega^2\,Y^*_{l_1m_1}Y_{l_2m_2}.
\end{align*}
If we rewrite the radial integral in terms of $r^*$, we then get\begin{align}
(\Psi_1,\Psi_2)
=& \frac{\omega_1+\omega_2}{4\pi\sqrt{\omega_1\omega_2}}e^{-i(\omega_2-\omega_1)t}
\int_0^\infty \text{d}r^*\rho_1^*(r)\rho_2(r) \nonumber\\
&\int_{S^2}\text{d}\Omega^2\,Y^*_{l_1m_1}Y_{l_2m_2}.
\end{align}

Now, the angular integral is just the usual inner product of spherical harmonics, which under the usual normalization is just $\delta_{l_1 l_2}\delta_{m_1 m_2}$. The radial integral is the usual $L^2$ inner product; however, since we know what the asymptotic behaviour of $\rho$ is, we can go further. Specifically, since $\rho_{\omega l} \rightarrow 2 \sin(\omega r^* + \zeta)$ for some phase $\zeta=\zeta(l,\omega)$, then the inner product is dominated by the behaviour at infinity (and thus we can ignore the part inside the shell), and so we can use standard arguments to find
\begin{align}
\int_0^\infty &\text{d}r^*\rho_1^*(r)\rho_2(r)\nonumber\\
\approx&\int_0^\infty \text{d}r^* (e^{i(\omega_1 r^*+\zeta_1)}-e^{-i(\omega_1 r^*+\zeta_1)})\nonumber\\
&\times(e^{-(i\omega_2 r^*+\zeta_2)}-e^{i(\omega_2 r^*+\zeta_2)})\nonumber\\
=&\int_0^\infty \text{d}r^* (e^{-i(\omega_2 r^*+\zeta_2-\omega_1 r^*-\zeta_1)}\nonumber\\
&-e^{-i(\omega_2 r^*+\zeta_2+\omega_1 r^*+\zeta_1)r^*}+h.c.)\nonumber\\
=&2\pi (\cos(\zeta_2-\zeta_1)\delta(\omega_2-\omega_1)\nonumber\\
&-\cos(\zeta_2+\zeta_1)\delta(\omega_2+\omega_1)),
\end{align}
where in the final line we use the fact that this integral is (half of) the Fourier transform of the Dirac delta distribution. Of course, since $\omega$ is positive, we can ignore the second term.

Substituting the values found for the two integrals finally gives us
\begin{align}
(\Psi_1,\Psi_2)
=& \frac{\omega_1+\omega_2}{4\pi\sqrt{\omega_1\omega_2}}e^{-i(\omega_2-\omega_1)t}\cos(\zeta_2-\zeta_1)\nonumber\\
&\times 2\pi \delta(\omega_2-\omega_1)\delta_{l_1 l_2}\delta_{m_1 m_2}\nonumber\\
=&\delta(\omega_2-\omega_1)\delta_{l_1 l_2}\delta_{m_1 m_2}
\end{align}
where we use the Dirac delta in the final line to simplify the prefactor.
\bibliography{shellscalarbib}

\end{document}